# Inhomogeneity identification by measuring magnetic quantum oscillations

Sang-Eon Lee[1], Myung-Hwa Jung[1,*]

[1]*Department of Physics, Sogang University, Seoul 04107, Korea*

This study explores the identification of sample inhomogeneity via magnetic quantum oscillations analysis in semimetal $NbSb_2$. By doping Bi and Cr, we obtained a homogeneous Bi-doped sample and an inhomogeneous Cr-doped sample, whose homogeneity was confirmed by comparing the magnetic quantum oscillation before and after grinding the samples. The magnetic quantum oscillations in the inhomogeneous sample exhibited a distinct phase shift and unusual field-dependent amplitude, believed to result from a non-uniform Fermi energy. The analysis of the magnetic quantum oscillations demonstrated that the homogeneous Bi-doped sample can be interpreted by the symmetric and Lorentzian effective Fermi energy distribution, while the inhomogeneous Cr-doped sample exhibited an asymmetric distribution, illustrating an unconventional violation of the Lifshitz-Kosevich formula. This research provides a novel method for identifying material inhomogeneity and mitigating potential misinterpretations of magnetic quantum oscillations' unusual phase, commonly seen as a nontrivial Berry phase indicator in topological materials studies.

**Keywords:** Inhomogeneity, non-uniform composition, de Haas-van Alphen effect

**Highlights**

• Pristine, Bi-, and Cr-doped NbSb$_2$ single crystals were synthesized by chemical vapor transport

• The de Haas-van Alphen (dHvA) effects of homogeneous Bi-doped sample and inhomogeneous Cr-doped sample were studied

• Unusual phase shift and unconventional field-dependency were observed in the dHvA oscillations of inhomogeneous Cr-doped sample

• From the analysis of dHvA oscillations, we found the non-uniform Fermi energy distribution of the inhomogeneous sample produces the unusual phase shift and unconventional field-dependency

# 1. Introduction

The inhomogeneity of materials has received much interest from researchers. The researchers have tried to get rid of inhomogeneities to improve material properties like carrier mobility [1–4], optical properties in solar cells [5–7], and topological properties of materials [8–11]. In contrast, researchers have studied products of inhomogeneity, including linear magnetoresistance [12–17], localization [18–20], topological properties [21,22] and thermoelectric properties [23]. In both perspectives, identifying the sample inhomogeneity is important for characterizing the sample and directing the synthesis procedure for the desired sample properties. There are various types of inhomogeneity, which can be non-uniform composition, dislocations, impurities, mosaic structures, and phase separations.

In the conducting alloys, the sample inhomogeneity produces a spatially non-uniform electromagnetic potential, therefore producing a spatially non-uniform Fermi energy distribution. Using the Fermi energy-detecting tool can be an effective way to investigate the inhomogeneity. The de Haas-van Alphen (dHvA) effect [24], which is the oscillatory phenomenon of magnetization as a function of the inverse magnetic field, $1/B$, is a typical tool to probe the Fermi surface [25–35], similar to its resistivity version of the Shubnikov-de Haas effect [36,37]. The variation of Fermi energy would lead to the variation of the frequency, $F$, of the dHvA oscillations by the famous Onsager's formula [38] $F = \hbar S_F / 2\pi e$, where $S_F$ is the extremal Fermi surface cross-section area perpendicular to the magnetic field direction. As a result, the spatially non-uniform Fermi energy causes dephasing effects, which can be caught by analyzing the phase and field-dependent amplitude of the dHvA oscillations. Based on this perspective, the inhomogeneity coming from impurities, dislocations, or mosaic structures has been studied with dHvA oscillations [39–45].

Here, we studied another prevailing inhomogeneity by the dHvA oscillations, the non-uniform composition. For the study, we used semimetal $NbSb_2$, whose dHvA oscillations have been clearly observed and well understood [28,46,47]. Furthermore, $NbSb_2$ is itself an interesting material with its large magnetoresistance, linear dispersion, and colossal Nernst power factor [28,46–48]. We artificially introduced inhomogeneity by doping Bi and Cr elements into $NbSb_2$. We focused on the homogeneous Bi-doped sample with spatially uniform Bi concentration and the inhomogeneous Cr-doped sample with spatially non-uniform Cr concentration. The inhomogeneities were confirmed by comparing dHvA oscillations before and after grinding the mm-size of the samples. The observed dHvA oscillations of inhomogeneous Cr-doped $NbSb_2$ were largely different from the expected form by the Lifshitz-Kosevich (LK) formula with an unusual phase shift and unconventional field dependency. This result is in contrast to the dHvA oscillations of the homogeneous Bi-doped sample, which well followed the LK formula. In addition, we obtained the effective Fermi energy distribution function by analyzing the dHvA oscillations. While the effective Fermi energy distribution function of the homogeneous sample was a Lorentziann function, the effective Fermi energy distribution function of the inhomogeneous sample is asymmetric and non-Lorentzian function.

This study demonstrates the non-uniform dopant composition can be analyzed by the dHvA oscillations and gives a potentially useful method for evaluating the inhomogeneity in conducting alloys. It is also notable that our study prevents possible misinterpretation of the unexpected phase shift from the inhomogeneity, which can be misunderstood as an indicator of the nontrivial Berry phase in the studies of topological materials [49–56].

## 2. Material and methods

2.1. Synthesis procedure

Pristine, Bi- and Cr-doped $NbSb_2$ were synthesized by chemical vapor transport (CVT). As a precursor, polycrystalline $NbSb_2$ was synthesized by solid-state reaction, where the stoichiometric mixture of high-purity Nb powder (99.99%, Alpha Aesar) and ground Sb shot (99.9999%, Alpha Aesar) was rapidly heated in a vacuum-sealed quartz tube up to 600°C, and slowly heated up to 700°C (2.5°C/h) to prevent rapid melting of Sb. Successively, the mixture was kept at 700°C for 36 hours. For the doped sample, the stoichiometries were adjusted to $Nb_{1-x}Cr_xBi_ySb_{2-y}$, where x = 0 and y = 0.2 for the Bi-doped sample and x = 0.1 and y = 0 for the Cr-doped samples with ground Bi needle (99.998%, Alpha Aesar), and ground Cr lump (99.99%, Alpha Aesar). We brought the polycrystalline samples to the end of the quart tube and set the temperature to 1000°C, and set the temperature of the other end of the quart tube to 900°C for 7 days. Several shiny and mm-size single crystals elongated along the *b*-axis were obtained at the low-temperature part of the quart tube. For the grinding experiment, we carefully ground samples with ultra-fine sandpapers (the average particle size ~ 10.3 μm), and ethyl-alcohol was used for lubrication.

2.2. Material characterization

The crystal structure of $NbSb_2$ is a monoclinic structure with C2/m space group, and it has a two-fold rotational symmetry about the *b*-axis. The crystal structure of our single crystals was confirmed by X-ray diffraction spectroscopy (XRD) (Rigaku DMAX 2500 diffractometer, Rigaku, Japan) with Cu Kα radiation (λ=1.5406 Å) operated at 40 kV and 15 mA. Obtained XRD data of pristine and doped samples were refined using Fullprof software, and the lattice parameters are the same up to second digits in the unit of Å (see Fig. S1 for the XRD data and Table. S1 – S4 for obtained parameters, respectively, in Supporting Information). The difference in the lattice parameters and Wyckoff positions of doped samples from those of pristine sample are listed in Table S5 and S6, respectively. The dopant composition ratios were identified using an electron probe X-ray microanalyzer (JXA-8530F, JEOL Ltd, Japan) with 15 keV and 20 nA electron current, which gives 0.40(2) and 0.05(3) atomic percent for Bi and Cr compositions. Note that the actual dopant compositions were much lower than the nominal compositions we put in since the excess of the doping elements was separated from the synthesized crystals in the CVT process.

2.3. Magnetic and electrical properties measurements

The magnetizations were measured by a commercial magnetic property measurement system (MPMS3 SQUID-VSM, Quantum Design) with the applied magnetic field up to 7 T. The magnetic field direction was along the crystal *b*-axis of single crystals in all magnetization data of this paper. The electrical resistances were measured by the standard 4-probe method using an electrical transport option of the MPMS3 in the temperature range of 2 ~ 300 K.

**3. Results and discussion**

3.1. The dopant distirubtion in Bi and Cr doped NbSb$_2$

We examined the sample inhomogeneity, which would be introduced by doping, by comparing the dHvA oscillations before and after grinding the millimeter-size samples. Here, we refer to the pristine, Bi-doped NbSb$_2$, and Cr-doped NbSb2 as pristine, Bi, and Cr, respectively. The dHvA oscillations were obtained from the field-dependent magnetizations measured at 2 K, after subtracting the background diamagnetic signals. The masses of Bi and Cr samples were reduced by the grinding process from 0.0367 and 0.0482 g to 0.0179 and 0.0136 g, respectively. We refer to these ground samples as Bi' and Cr', respectively. The dHvA oscillations of both unground and ground samples were compared in Fig. 1(a, b). While there is minimal difference between the dHvA oscillations of Bi and Bi', there is a notable difference between those of Cr and Cr'. These results indicate that there is significant millimeter-range inhomogeneity in the Cr sample. For the Bi sample, it appears that Bi dopants create a short-range local electric potential variation but do not cause long-range inhomogeneity due to the homogeneous distribution of the dopants. In Fig. 1(c, d), we illustrate the distributions of the dopants compared to the cyclotron orbits to elucidate the relationship between homogeneity and dHvA oscillations. For the Bi sample, the surrounding areas near the cyclotron orbits are similar regardless of whether the regions were ground out (the outside of the red rectangle) or preserved after grinding (the inside of the red rectangle), producing similar dHvA oscillations before and after grinding. In contrast, for the Cr sample, the surrounding areas of the cyclotron orbits differ between the two regions (inside and outside of the red rectangle), suggesting the different dHvA oscillations generated between the two regions. This leads to the discrepancy in the dHvA oscillations before and after grinding.

We analyzed the dHvA effects in detail. The fast Fourier transform (FFT) was conducted to investigate the frequencies of the dHvA oscillations. Fig. 2(a) shows the FFT spectra of pristine, Bi, and Cr at 2 K. Several peaks are observed within the range of 100 ~ 1000 T, which is consistent with the previous studies about $NbSb_2$ [28,46,47]. The peaks were indexed as ξ (~ 63 T), ζ (~ 199 T), α (~ 383 T), β (~ 705 T), and 2β (~ 1409 T). Notably, ξ peak was only observed in the Cr sample, which can come from an additional phase produced by the phase separation in the Cr sample. We analyzed β peak selectively for the comparison study since only β peak was clearly observed in all samples. For that, we used the FFT bandpass filter to separate β oscillations. We analyzed the β oscillations with the LK formula, where the fundamental harmonics of dHvA oscillations for the three-dimensional ellipsoidal Fermi surface, $\Delta M$ is expressed as follows [25,26,28].

$$\Delta M \propto -\sqrt{B} \ |\cos\lambda|\ R_T R_D \sin(2\pi F / B - \pi + \Theta - \pi/4), \qquad (1)$$

where $\lambda$ is the phase from the magnetic properties of the Fermi surface, $R_T = (2\pi^2 m_c k_B T/e\hbar B)/\sinh(2\pi^2 m_c k_B T/e\hbar B)$ is the temperature-reduction factor, $R_D = \exp(-2\pi^2 m_c k_B T_D/e\hbar B)$ is the Dingle reduction factor, $\Theta = \pi\{1 - \text{sign}(\cos\lambda)\}/2$, $m_c$ is the cyclotron mass, and $T_D$ is the Dingle temperature, which is proportional to the scattering rate. The amplitude factor $\sqrt{B}$ and the phase $-\pi/4$ come from the shape of the three-dimensional ellipsoidal Fermi surface. We first analyzed the temperature-dependent amplitudes of the β oscillations by fitting them with $R_T$ to verify how the doping changes $m_c$. Fig. 2(b) shows the temperature-dependent amplitudes measured at $B = 6.5$ T and the fitted curve by $R_T$. We obtained $m_c = 0.51$ for all samples, indicating that doping has no effect on the cyclotron masses.

Successively, we analyzed field-dependent amplitudes obtained at 2 K to investigate $T_D$. To eliminate other effects on the field-dependency of amplitudes except for $R_D$, we defined $D = \text{Amplitude}/ R_T \sqrt{B}$, and conducted Dingle plots ($\ln D - B^{-1}$) in Fig. 2(c). Interestingly, the Dingle plot of Cr shows a nonlinear shape, while those of the others show linear shapes. For the samples with linear Dingle plots, we obtained $T_D$ = 0.16, and 1.3K for pristine, and Bi samples by fitting the linear slopes with $-2\pi^2 m_c k_B T_D/e\hbar B$. Increased $T_D$ by Bi doping shows increased scattering events by Bi doping. The increased scattering events by Bi and Cr dopants were also verified by the temperature-dependent resistivity, which shows the residual resistivities at 2 K are 0.08, 1, and 5 μΩcm and the residual resistivity ratios, $\rho$(300 K)/$\rho$(2 K), are 790, 64 and 13 for pristine, Bi, and Cr, respectively (see Fig. S2 in Supporting Information). It is notable that Cr dopants act more effectively as scattering centers (increase the low-temperature resistivity) than Bi dopants do. However, the electron probe X-ray microanalyzer results show a higher Bi composition (~0.40(2) atomic percent) than Cr composition (~0.05(3) atomic percent), although the Cr composition is not very accurate. These results suggest that each Cr dopant gives stronger electrostatic perturbation than each Bi dopant does. This fact may be explained by much similar chemical properties between Sb and Bi (both have 3 electrons in p orbitals and 2 electrons in s orbitals as valence electrons) than Nb and Cr (Nb has 4 electrons in d orbitals and one electron in s orbital as valence electrons, but Cr have one more electron in d orbital).

The phases of β oscillations were also analyzed. We assigned $N = n + 1/4$ ($n + 3/4$) for $B_{min(max)}$ and plotted the Landau fan diagrams (1/ $B_{min(max)}$ – N), where $n$ is a positive integer. In the process, sets of $n$ are chosen to make $N$-intercept, $\gamma$, in a range of –0.5 ~ 0, where $\gamma$ indicates normalized phase $\phi/2\pi$ (mod 1). Fig. 2(d) shows the Ladan fan diagrams. We obtained $\gamma$ = –0.11, –0.1, and –0.46 for pristine, Bi, and Cr, respectively. There is a large deviation between the theoretical value –0.125 with $\Theta = \pi$ [28], and the phase of the Cr sample. We also obtained $F$ = 704.6, 708.9, and 714.6 from the slopes of the Landau fan diagrams for pristine, Bi, and Cr, respectively, indicating that the doping slightly increases the Fermi surface. But the overall sizes are unaltered, like in the case of $m_c$.

We used the effective Fermi energy distribution, D($\mu$), for the Fermi energy, $\mu$, to explain the nonlinear Dingle plot and unusual phase shift in the inhomogeneous Cr sample. In this model, observed magnetic quantum oscillations are effectively viewed as a superposition of all oscillations from different $\mu$, whose distribution follows D($\mu$). This idea has been used to account for the reduction factors from the finite temperature and relaxation time effects [25]. Since only the relative energy difference between the Fermi energy and the energy of occupied states matters in the magnetic quantum oscillations, the occupation distribution (Fermi-Dirac distributiontion by the finite temperature) and the probability distribution of state energy (the Lorentzian distribution for the energy uncertainty by the finite relaxation time) can be equally viewed as a corresponding effective Fermi energy distribution. In this paper, we separated the temperature effect from D($\mu$), where the dHvA oscillation is given by $\Delta M \propto -R_T \sqrt{B} \int_{-\infty}^{\infty} D(\mu) \sin[2\pi F(\mu)/B - \pi/4] d\mu$. For the homogenous samples, D($\mu$) become Lorentzian probability distribution from the finite relaxation time and give a linear Dingle plot, which was suggested by Dingle [43]. In that case, the full width at half maximum (FWHM) of the Lorentzian function $\Delta\mu$ determine the Dingle temperature as $T_D = \Delta\mu/\pi k_B$. For the inhomogenous samples, the spatial difference of the Fermi energy should be included in D($\mu$) with the probability distribution, producing a deviation from the linear Dingle plot. Furthermore, the spatially inhomogeneous Fermi energy can make D($\mu$) asymmetric, giving the unusual phase shift. We interpreted the dHvA oscillations of the Cr sample as a result of asymmetric and non-Lorentzian D($\mu$). As a trial solution for D($\mu$) of the Cr sample, we adopted the multi-Lorentzian function, $D(\mu) = \sum_i A_i L(\mu; \mu_{0i}, \Delta\mu_i) / \sum_i A_i$, where $L(\mu; \mu_0, \Delta\mu)$ is the Lorentzian function with the center $\mu_0$ and its FWHM, $\Delta\mu$. For the multi-Lorentzian distribution function, $\Delta M$ divided by $R_T \sqrt{B}$, $D_{osc}$, is given by

$$D_{osc} \propto -\sum_i A_i R_D(\Delta\mu_i) \sin[2\pi F(\mu_{0i})/B - \pi - \pi/4] / \sum_i A_i. \qquad (2)$$

We fitted $D_{osc}$ of the Cr sample by Eq. (2) with the three Lorentzian functions. Fig. 3(a) shows $D_{osc}$ vs. $B^{-1}$ of the Cr sample for β oscillations at 2 K and fitted curve. The fitted curve captures the overall shape of the data despite the data is not perfectly fitted, which may be attributed to our simple trial solution and the small dHvA oscillations signal of the Cr sample ~ 1 μemu (easily affected by the noise). In Fig. 3(b), we separately plotted the three oscillations constituting the fitted curve. It is in contrast to $D_{osc}$ of the Bi sample, which is well fitted by single oscillations, which is shown in Fig. 3(c). From the fitting, we obtained $A_i / \sum_i A_i$ = 0.4322, 0.0755, 0.4923, $F(\mu_{0i})$ = 710.9, 713.3, 716.4 T, and $\Delta\mu$ = 1.55, 0.88, 1.43 meV for Cr sample. For the Bi sample, we obtained $F$ = 708.7 T, and $\Delta\mu$ = 0.40 meV.

We also visualized D($\mu$) in Fig. 4. We set $\mu = 0$ for the center of D($\mu$) of the pristine sample. The effective Fermi energy distribution function is shown as a function of the Fermi energy deviation, $\delta\mu$, from $\mu = 0$. The center difference of D($\mu$) of doped samples was evaluated by the relation $\delta\mu \approx e\hbar\Delta F/m_c$, where $\Delta F$ is the frequency variation from the $F$ of pristine and $m_c = 0.51 m_e$. The very sharp peak of pristine with $\Delta\mu$ = 43 μeV indicates the well-defined Fermi energy and its high crystal quality. For the Bi sample, while its peak is broader than that of pristine, the peak is symmetric and follows the Lorentzian form with $\Delta\mu$ = 0.40 meV following Dingle's original idea. The peak of the inhomogeneous Cr sample has a broader HWHM = 1.7 meV than that of Bi, asymmetric and non-Lorentzian, producing the phase deviation and the nonlinear Dingle plot.

We also analyzed the dHvA oscillations of the Cr' sample (see Fig. S3 in Supporting Information), and interestingly, it seems that the inhomogeneity was relaxed by grinding. We obtained $F$ = 717.1 T for β oscillations of the Cr' sample, which is larger than that of the Cr sample. Given that Cr dopants increase the cross-section area of the Fermi surface for β oscillations, we deduced that there is more Cr concentration in the Cr' sample than the Cr sample. This result also indicates the inhomogeneous Cr composition in the Cr sample. The peak ξ observed in the Cr sample was not observed in the Cr' sample, suggesting the peak ξ is a result of the additional phase and ground out in the Cr' sample. The Dingle plot of the Cr' sample is much more linear than that of the Cr sample, and we obtained $\gamma$ = –0.07, which is quite close to the expected value of –0.125. This implies a much weaker inhomogeneity of the Cr' sample and that the inhomogeneity was relaxed by grinding.

Finally, we excluded other possible origins of the nonlinear Dingle plot and the unusual phase shift of Cr. First, the field-dependent magnetic susceptibilities can be attributed to the nonlinear Dingle plot, but we excluded the possibility by confirming the little field dependency in the magnetic susceptibilities (see Fig. S4 in Supporting Information). Second, the nonlinear Dinge plot and unusual phase shift can be produced by the magnetic nature of Cr impurities. For the case, the s-d coupling [57], spin-dependent scattering, and the exchange field [58–60] can generate the nonlinear Dingle plot and unusual phase shift. It was checked by measuring temperature-dependent magnetic susceptibilities with the applied magnetic field, $B$ = 1 T (see Fig. S5 in Supporting Information). The susceptibilities are diamagnetic and almost temperature-independent, $\chi(2\ \text{K})/|\chi(300\ \text{K})| \sim 0.9$, for all samples, which shows the nonmagnetic nature of dopants. Based on the results, we excluded the case of magnetic impurities for the Cr sample.

## 4. Conclusion

In this study, we introduced artificial inhomogeneity into $NbSb_2$ by doping Bi and Cr elements and evaluating the inhomogeneity by measuring the dHvA oscillations. We found the non-uniform Cr composition in Cr-doped $NbSb_2$ from the large difference of the dHvA oscillations between unground and ground Cr-doped $NbSb_2$, suggesting the mm-scale range of the inhomogeneity. We demonstrated that the inhomogeneity produces the nonlinear Dingle plot and the unusual phase deviation of the dHvA oscillations. On the other hand, Bi dopants are homogeneously distributed in Bi-doped $NbSb_2$, and such unconventional dHvA oscillations were not observed. Our study highlights how the inhomogeneity originating from non-uniform dopants concentration can be systematically studied by the dHvA oscillations. We also suggest that it is valuable to conduct a similar study on two-dimensional materials or thin film. The thin thickness of those samples would make identifying and imaging the dopant distribution more feasible. It would also be worthful to study how inhomogeneity is produced in the CVT process for the specific system and how the inhomogeneity can be controlled, which remains for future study topics.

**CRediT authorship contribution statement**

**S-.E. Lee:** Conceptualization, Methodology, Formal analysis, Investigation, Writing – Original draft **M-.H. Jung:** Conceptualization, Writing – Review & Editing, Supervision, Project administration

**Data availability**

Data will be made available on request.

**Declaration of Competing Interest**

The authors declare that there is no competing interest.

**Acknowledgments**

This work was supported by a National Research Foundation of Korea (NRF) grant (No. 2020R1A2C3008044).

**Appendix A. Supporting information**

Supplementary data associated with this article can be found in the online version at "".

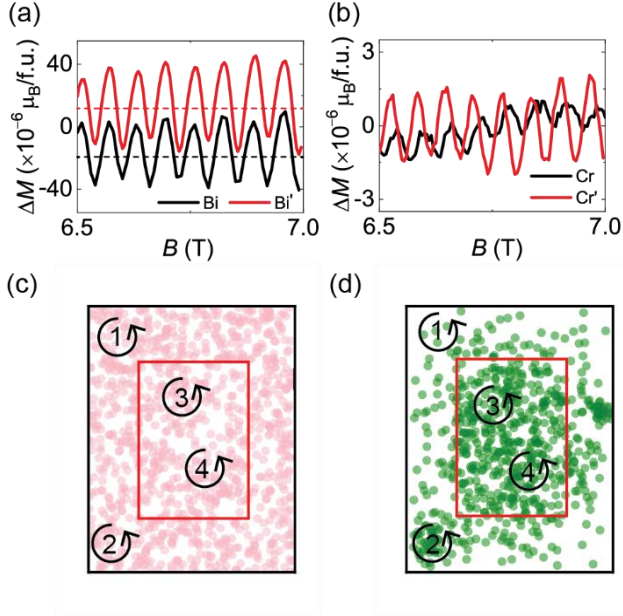

Fig. 1. Comparison of the dHvA oscillations before and after grinding mm-size samples. The dHvA oscillations were measured at 2 K. The ground samples are indicated by a prime. (a) The dHvA oscillations of Bi-doped NbSb$_2$ before and after grinding. We vertically shifted Bi and Bi' data for clarity (the dashed lines show the shifted $\Delta M = 0$ lines). There is no significant difference between the two, indicating homogeneity. (b) The dHvA oscillations of Cr-doped NbSb$_2$ before and after grinding show a significant difference, indicating mm-size range inhomogeneity in the Cr-doped sample. (c), (d) Illustrations of dopant distributions. The outer black rectangle and inner red rectangle display the sample boundaries before and after grinding, respectively. The circular arrows represent the cyclotron orbits. The Bi dopants are depicted as pink circles in (d). The homogeneous distribution of Bi dopants results in similar dHvA oscillations before and after grinding due to the similar surrounding areas of the cyclotron orbits. In contrast, the Cr dopants, illustrated as green circles in (d), have an inhomogeneous distribution. Consequently, the cyclotron orbits on the outer side (regions 1 and 2) differ from those on the inner side (regions 3 and 4), causing variations in the dHvA oscillations before and after grinding. In the illustrations, the size or cyclotron orbits and dopants were exaggerated compared to the sample size for visality.

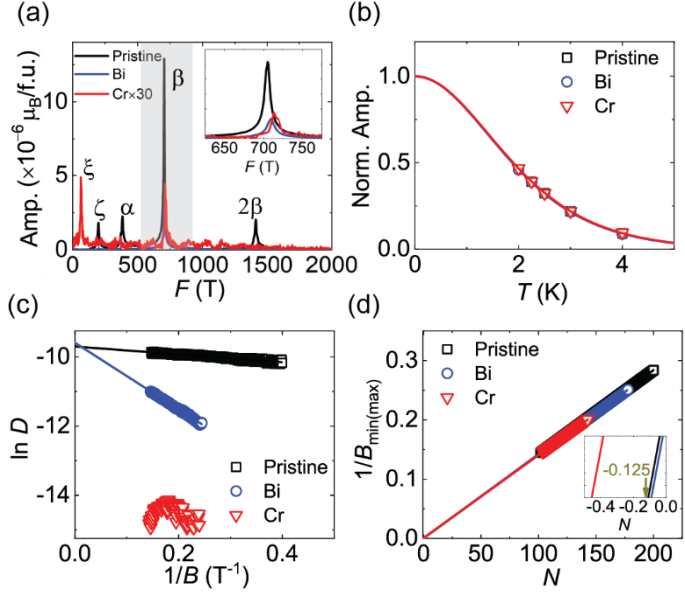

Fig. 1. dHvA oscillations. (a) The FFT spectra of the dHvA oscillations measured at 2 K. The amplitude of Cr is magnified by 30 times for visibility. The grey area indicates the FFT bandpass filter range to separate β oscillations. The inset shows the magnified FFT spectra focusing on β oscillations. (b) The temperature-dependent amplitudes of β oscillations. The amplitudes are normalized to have value unity at 0 K. The overlapped curves indicate a similar $m_c = 0.51 m_e$. (c) The Dingle plot (ln $D$ vs. $1/B$) of β oscillations at 2 K, where $D$ is defined as $D =$ Amplitdue/$R_T \sqrt{B}$. The nonlinear Dingle plot of Cr sample indicates the non-Lorentzian Fermi energy distribution function by the inhomogeneity. From the slopes of pristine and Bi, $T_D = 0.16$, and 1.3K are evaluated, respectively. (d) The Landau fan diagrams of β oscillations at 2 K. The inset shows the magnified Landau fan diagrams focusing on the intercepts. While the intercepts of pristine and Bi samples are close to the expected value −0.125, the intercept of Cr is far from −0.125, indicating the asymmetric Fermi energy distribution by the inhomogeneity.

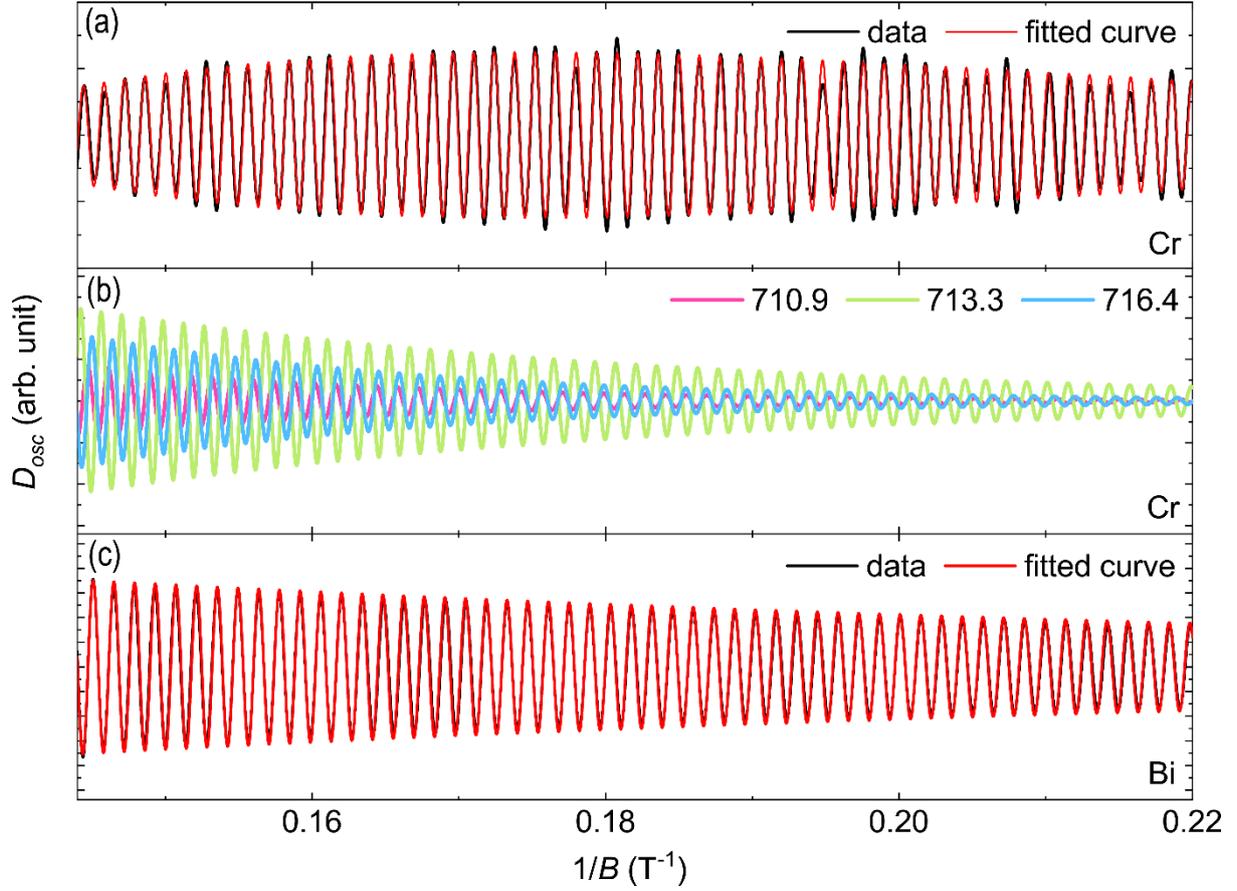

Fig. 3. β oscillations and fitted curve with Eq. (2). (a) The β oscillations of Cr sample with reduced amplitude, $D_{osc}$ (the black line). $D_{osc}$ is defined as $D_{osc} = \Delta M / R_T \sqrt{B}$. The red line is fitted curve with Eq. (2). The oscillations are well-fitted with three oscillations with parameters $A_i / \sum_i A_i$ = 0.4322, 0.0755, 0.4923, $F(\mu_{0i})$ = 710.9, 713.3, 716.4 T, and $\Delta\mu$ = 1.55, 8.84, and 1.43 meV. (b) The fitted curve of (a) is shown as three separated exponentially damped oscillations. The legend indicate the frequencies of oscillations in T unit. (c) $D_{osc}$ of Bi sample (the black line). The red line is fitted curve with Eq. (2). The oscillations is well-fitted by a single exponentially damped oscillations. We obtained $F$ = 708.7, and $\Delta\mu$ = 0.40 meV from the fitting.

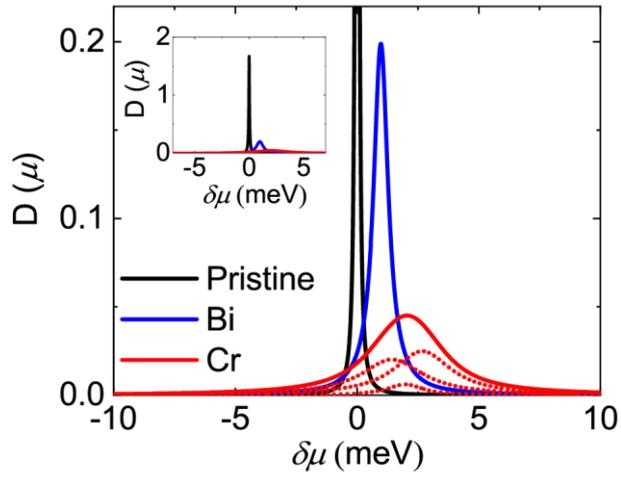

Fig. 4. Magnified effective Fermi energy distribution. The dotted red curves show that the asymmetric and non-Lorentzian Fermi energy distribution of the Cr sample can be expressed by the sum of three Lorentzian effective Fermi energy distributions. $\delta\mu$ indicates the relative energy from the Fermi energy of pristine. The inset shows unmagnified effective Fermi energy distribution, emphasizing the sharp effective Fermi energy distribution of pristine.